\documentclass[journal,twoside,web]{ieeecolor}
\usepackage{generic}
\usepackage{cite}
\usepackage{amsmath,amssymb,amsfonts}
\usepackage{algorithmic}
\usepackage{graphicx}
\usepackage{textcomp}
\usepackage{orcidlink}
\usepackage[none]{hyphenat}
\def\BibTeX{{\rm B\kern-.05em{\sc i\kern-.025em b}\kern-.08em
    T\kern-.1667em\lower.7ex\hbox{E}\kern-.125emX}}
\begin{document}

\title{On the $\mathrm{In_{x}Ga_{1-x}As}$ channel noise in\\ 
InP HEMTs from 4 K to 300 K}

\author{Junjie Li \orcidlink{0000-0002-0081-3365}, \and
Justin H. Chen \orcidlink{0000-0001-9745-7055}, \and
Austin J. Minnich \orcidlink{0000-0002-9671-9540}, \and
and Jan Grahn \orcidlink{0000-0002-5439-870X}
\thanks{Junjie Li was with the Department of Microtechnology and Nanoscience, Chalmers University of Technology, SE-41296 Gothenburg, Sweden. She is now with the Chalmers Next Labs AB, WACQT Quantum Technology Testbed, SE-41296 Gothenburg, Sweden (e-mail: junjie.li@chalmersnextlabs.se)}.
\thanks{Jan Grahn is with the Department of Microtechnology and Nanoscience, Chalmers University of Technology, SE-41296 Gothenburg, Sweden (e-mail: jan.grahn@chalmers.se)}
\thanks{Justin H. Chen and Austin J. Minnich are with the Division of Engineering and Applied Science, California Institute of Technology, Pasadena, CA 91125 USA (e-mail: jchen10@caltech.edu; aminnich@caltech.edu)}}
\maketitle

\begin{abstract}
The InP high-electron-mobility transistor (HEMT) is indispensable for low-noise amplifiers (LNAs) in radio astronomy and quantum computing. The composition of the $\mathrm{In_{x}Ga_{1-x}As}$ channel in the InP HEMT is known to influence the LNA noise performance. However, the various physical mechanisms responsible for noise generation are not fully characterized and understood. Here, we investigate the $\mathrm{In_{x}Ga_{1-x}As}$ channel noise from 4 K to 300 K for 100-nm gate-length InP HEMTs with channel indium content of 53\%, 60\% and 70\%. Channel noise was quantified by extracting the equivalent drain noise temperature $\mathit{T}_{d}$ using both on-wafer and LNA-based measurements, covering 40-300 K and 4-40 K, respectively. The 60\% indium channel InP HEMT exhibited the lowest channel noise across the full temperature range. The $\mathit{T}_{d}$ extracted from on-wafer characterization was found to obey a parabolic temperature dependence which predicted the $\mathit{T}_{d}$ at 4 K for all InP HEMTs in good agreement with LNA-based measurements. By expressing the channel noise as the sum of one thermal and one excess noise term, it was found that the former increased linearly with ambient temperature and dominated at 300 K. The channel noise at 4 K was determined by the excess noise term and exhibited a non-monotonic dependence on the channel indium content in the InP HEMT. The results suggest that the excess noise in the $\mathrm{In_{x}Ga_{1-x}As}$ channel originates not only from temperature-independent shot noise but also from impact ionization and real-space transfer noise.

\end{abstract}


\section{Introduction}
\label{sec:introduction}
The InP high-electron-mobility transistor (HEMT) is used to design cryogenic low-noise amplifiers (LNAs) with the highest gain and lowest noise from microwave up to mm-wave frequencies \cite{Bardin2021}. Such LNAs constitute critical front-end components in radio astronomy \cite{Bryerton2013} and superconducting quantum computing readout systems \cite{Hornibrook2015}. The InP HEMT is the preferred active device for cryogenic LNAs because the low electron effective mass and high peak electron velocity of the $\mathrm{In_{x}Ga_{1-x}As}$ channel \cite{HelenaTED} in the transistor at low temperature enable a high current-gain cut-off frequency already at a drain current density $I_{d}$ of a few tens of mA/mm \cite{Cha2023}. This results in a minimum noise temperature $T_{min}$ of approximately 1 K at 6 GHz for the cryogenic InP HEMT, the lowest among all transistor technologies \cite{Schleeh2012}. A state-of-the-art 4–8 GHz hybrid InP HEMT cryogenic LNAs was reported more than a decade ago with an average noise temperature $T_{e,avg}$ of 1.6 K and an average gain of 44 dB at a total power dissipation of 4.2 mW \cite{Schleeh2012}. Recently, a similar design was demonstrated with $T_{e,avg}$ =1.4 K at 2.1 mW \cite{Rebelo2026}. The high gain at low drain current density also makes the cryogenic InP HEMT LNA uniquely suited for ultra-low-power operation, exhibiting $T_{e,avg}$ =3.3 K at a power of only 108 $\mu$W while maintaining an average gain of 21 dB for the 4-8 GHz band \cite{Li2024}. The outstanding low-noise properties of the InP HEMT have also been exploited in wideband monolithic microwave integrated circuits for cryogenic applications; A 0.3–14 GHz LNA with $T_{e,avg}$ = 3.5 K \cite{Cha2018} and a 70-96 GHz LNA with a minimum noise temperature of 22 K at 85 GHz \cite{Bryerton2009}. Using a similar $\mathrm{In_{x}Ga_{1-x}As}$ channel structure, 8-18 GHz GaAs mHEMT LNAs were also reported with $T_{e,avg}$ = 4.2 K \cite{Heinz2021}, and $T_{e,avg}$ = 11.2 K over 16 -50 GHz \cite{Heinz2026}. 

The noise performance of the cryogenic LNA has been shown to be highly sensitive to the channel indium content of the InP HEMT \cite{ChaIMS}. It was previously reported that the noise temperature for InP HEMTs was lower for 65\% indium content channel than for 80\% channel at 4 K \cite{Cha2023}. Furthermore, it was also disclosed in \cite{Ruiz2019} and \cite{Wang2022} that increasing the channel indium content of the InP HEMT resulted in higher noise at 300 K. The generation of noise in the InP HEMT is dominated by the properties of the $\mathrm{In_{x}Ga_{1-x}As}$ channel and is due to several physical mechanisms which are not fully understood. Since thermal noise alone can not describe the channel noise in cryogenic field-effect transistors \cite{Smit2014,Marian2005}, alternative noise sources have been proposed for the InP HEMT, $e.g.$ suppressed shot noise \cite{Marian2017}, real-space transfer (RST) noise\cite{Tomi2022,Li2022}, and impact ionization noise \cite{Ruiz2019,RuizIEDM}. In \cite{Li2024}, we reported an optimum channel indium content of 60\% for the lowest noise in the 4-8 GHz (C-band) InP HEMT LNA at 4 K. It was suggested that this was due to a combination of thermal noise and RST excess noise in the $\mathrm{In_{x}Ga_{1-x}As}$ channel. However, the noise characterization was limited to 4 K, leaving the temperature dependence of the channel noise and its underlying mechanisms unresolved.

In this paper, we present channel noise measurements for 100-nm gate-length InP HEMTs based on different indium channels for the full temperature range from 4 K to 300 K. Contributions from thermal and excess noise in the channel have been extracted using both on-wafer and LNA-based measurements. Variations in channel noise with different channel indium contents and temperatures were used to identify the physical mechanisms governing the noise in the InP HEMT.

\section{Experiment}
\begin{figure}[h!]
\centerline{\includegraphics[width=1\columnwidth]{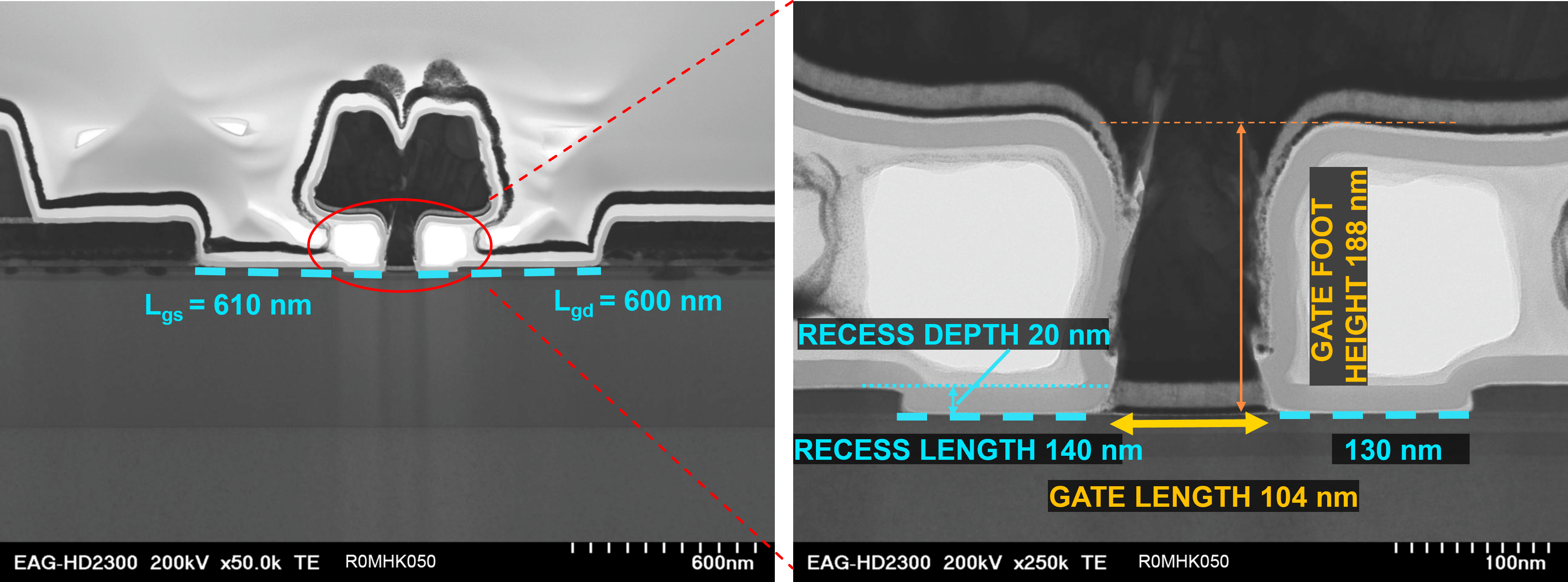}}
\caption{Cross-sectional STEM images of the fabricated InP HEMT. Left: overview of the T-shape gate showing source–gate spacing $L_{gs}$ and drain–gate spacing $L_{gd}$; the red circle marks the gate region magnified at right. Right: view of the recessed gate.}
\label{SEM}
\end{figure}

The InP HEMTs used for noise characterization were fabricated simultaneously in a 100-nm gate-length process \cite{Cha2018}. The cross-section of the T-shape gate region is shown in Fig. \ref{SEM} including dimensions, with source-gate spacing $L_{gs}$ = 610 nm, drain–gate spacing $L_{gd}$ = 600 nm, recess length of 140 nm and 130 nm for left and right, respectively. Apart from the channel, the InP HEMTs had identical epitaxial structures, see details in \cite{Li2024}. The channel was a homogeneous $\mathrm{In_{x}Ga_{1-x}As}$ layer with x = 0.53, 0.60, or 0.70, and a thickness of 15, 20 and 10 nm, respectively. Under low-noise transistor operation, the two-dimensional electron gas (2DEG) will be closely located to the barrier/channel interface as seen from simulations \cite{Cha2023, HelenaTED}. As a result, the differences in channel thickness are not expected to affect the channel noise in the InP HEMT\cite{Li2024}. 

\begin{figure}[h!]
\centerline{\includegraphics[width=1.1\columnwidth]{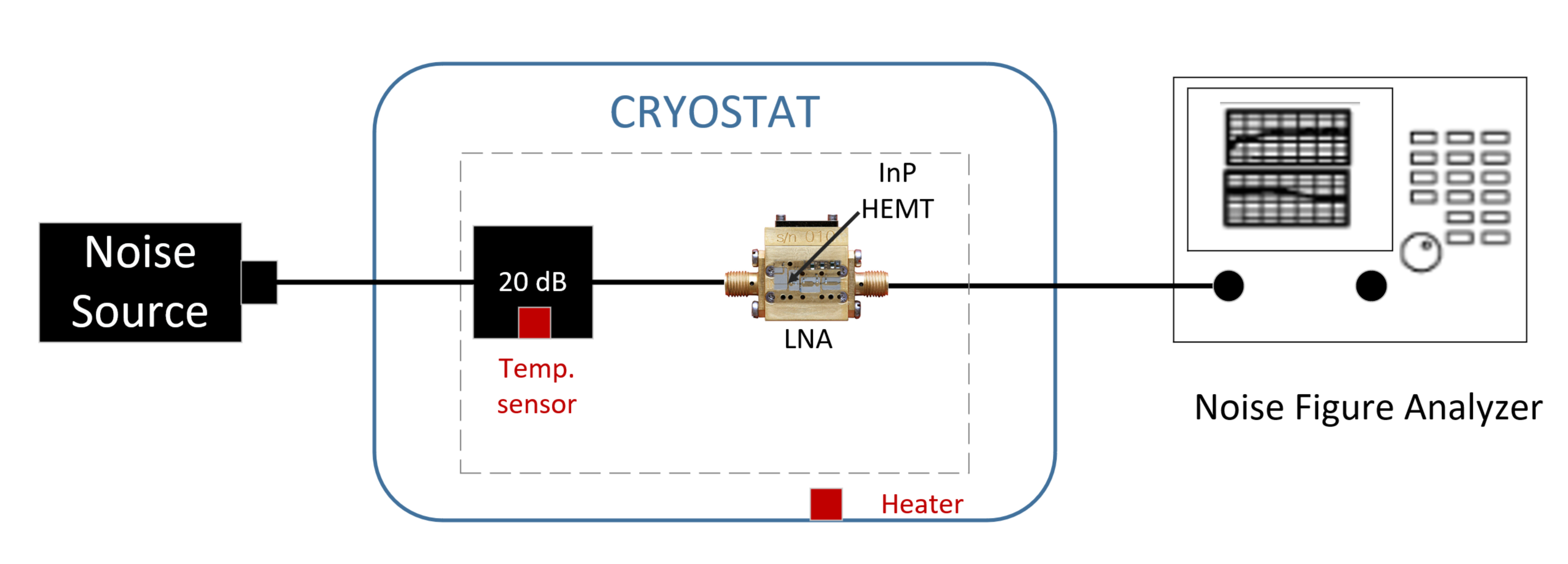}}
\caption{Schematic of the cryogenic noise measurement setup using an LNA module equipped with the InP HEMTs. The LNA is cooled to a temperature between 4 and 40 K inside a cryostat.}
\label{LNAmeas}
\end{figure}

\begin{figure}[h!]
\centerline{\includegraphics[width=1\columnwidth]{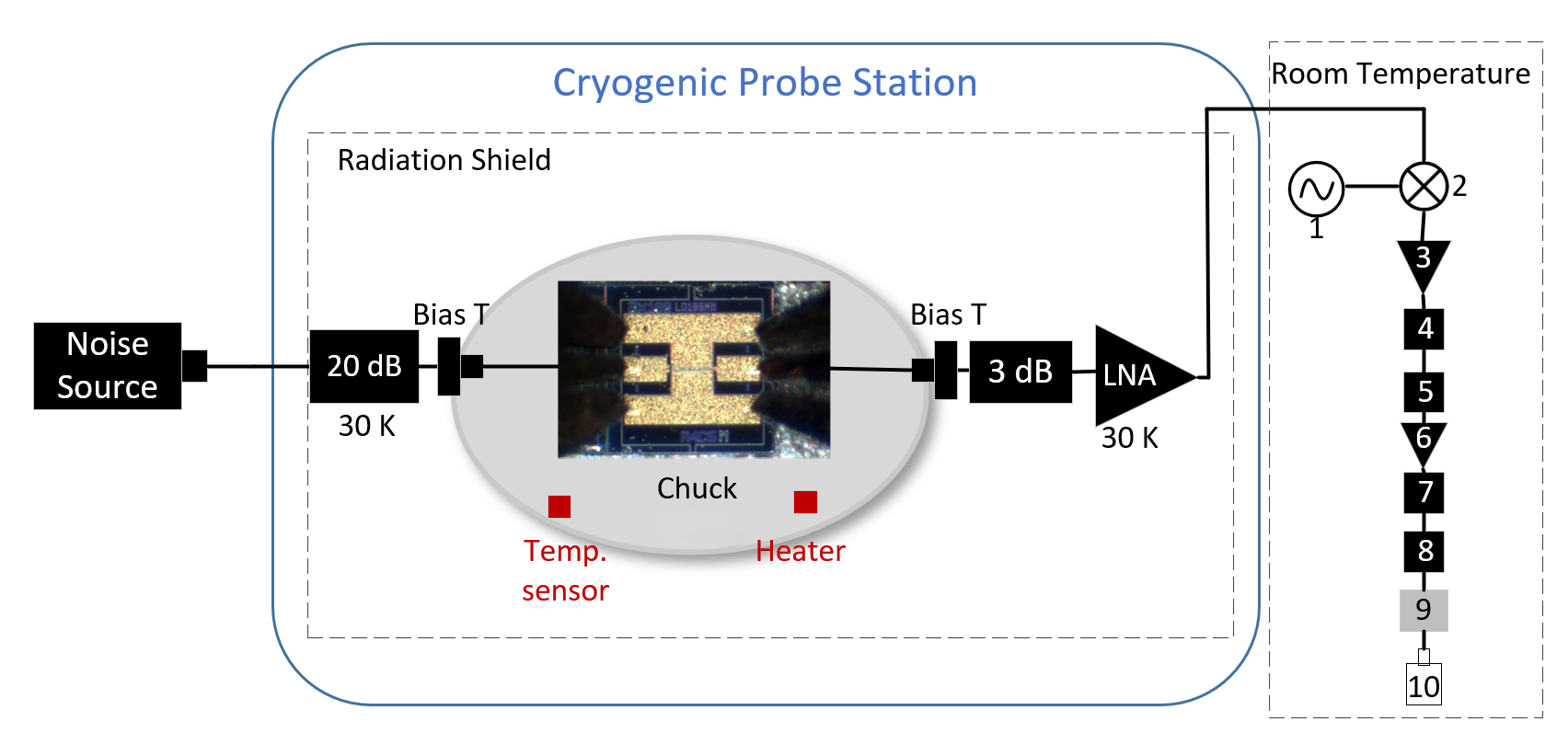}}
\caption{Schematic of the on-wafer noise measurement setup in a cryogenic probe station. The InP HEMT is probed on-wafer at temperatures from 40 to 300 K. The room-temperature measurement chain consists of (1) HMC-T2100 local oscillator, (2) M1-0220-P mixer, (3)–(4) low-noise amplification and attenuation, (5)–(8) bandpass filters, (9) variable attenuator, and (10) U8481A power sensor.}
\label{onwafermeas}
\end{figure}

Depending on the measurement ambient temperature $T_{a}$, two different setups were employed for estimating the InP HEMT noise from 4 K to 300 K. As a result, a high measurement accuracy was obtained in the full temperature range. At 40 K $>$ $T_{a}  \ge$ 4 K, a hybrid three-stage C-band LNA equipped with the InP HEMTs was used as shown in Fig.\ref{LNAmeas} \cite{Wadefalk2003,Gallego1990}. Provided the LNA gain was higher than 40 dB, the LNA $T_{e,avg}$ at 4 K was estimated to have an accuracy of 20\% \cite{Li2024}. Since uncertainty in LNA noise measurement increased with temperature \cite{Cano2010}, an on-wafer setup was used for estimating transistor noise at  $T_{a}  \ge$ 40 K. For this purpose, a cryogenic probe station was used as illustrated in Fig.\ref{onwafermeas} with a temperature sensor positioned in close proximity to the probed InP HEMT \cite{Russell2012}. The accuracy of the system noise temperature $T_{50}$ (with a 50-Ohm generator impedance) in the frequency range 4.0-7.2 GHz was estimated to 20\% above 40 K \cite{Beka2022}. For each channel indium content, five different InP HEMTs were measured. For each temperature and device, an average noise value was calculated from at least three independent noise measurements.

\section{DC and RF characterization}

\begin{figure}[!t]
\centering
\includegraphics[width=\columnwidth]{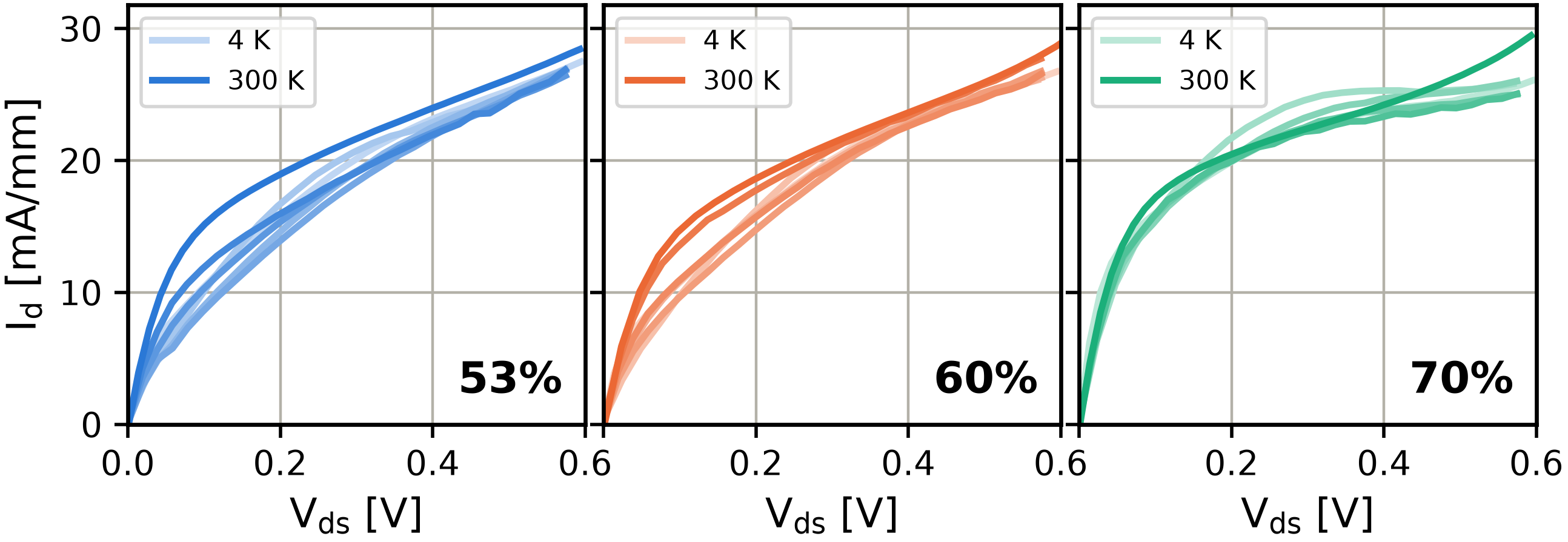}
\caption{Measured output characteristics $I_d$-$V_{ds}$ of the 53\%, 60\% and 70\% indium channel InP HEMTs. Trace shading indicates the ambient temperature from
$T_a = 4$~K (lightest), 20 K, 40 K, 80 K, 100 K, 200 K to 300~K (darkest).}
\label{dc}
\end{figure}

Fig.~\ref{dc} shows the measured $I_d$--$V_{ds}$ for the three InP HEMTs from $T_a = 4$~K to 300~K. For each $T_a$, the gate voltage $V_{gs}$ was tuned to give the $I_{d}$ aimed for the noise measurements. The dependence on channel indium content is visible in the shape of the output characteristics. From drain voltage $V_{ds} \approx 0.5$~V, the 70\% indium channel InP HEMTs exhibits a minor kink effect not seen for 53\% and 60\% devices, in agreement with Ref.\cite{Cha2023}, associated with impact ionization and surface traps \cite{HelenaTED}.

\begin{figure}[!t]
\centering
\includegraphics[width=\columnwidth]{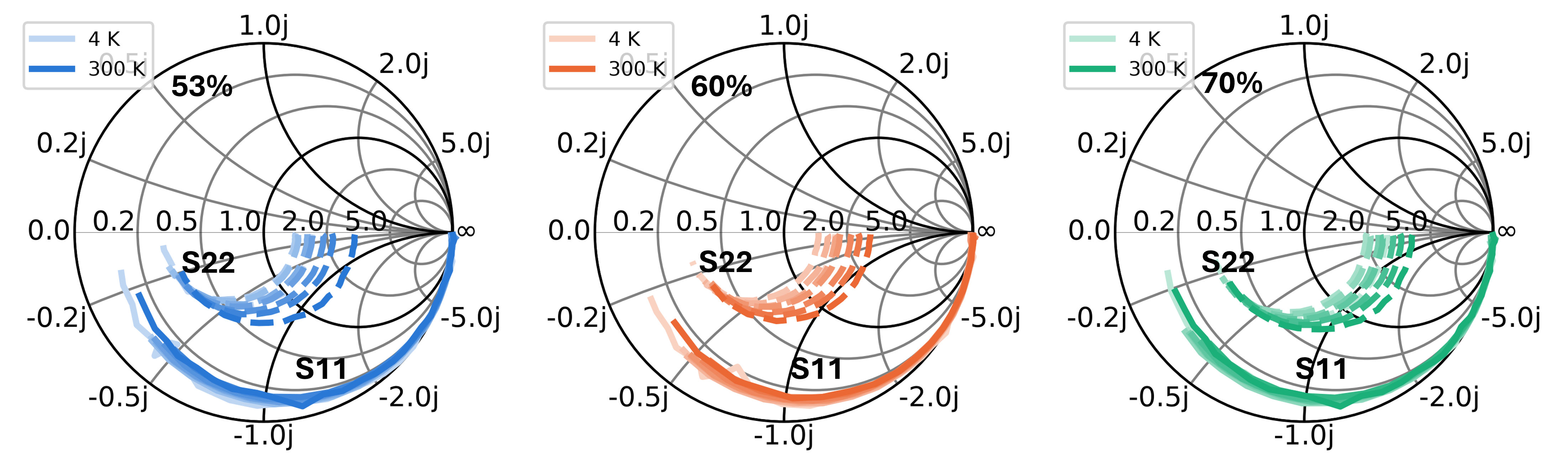}
\caption{Measured input and output reflection coefficients $S_{11}$ and
$S_{22}$ for 53\%, 60\% and 70\% indium channel InP HEMTs at
$V_{ds} = 0.5~\rm V$ and $I_d = 25~\rm mA/mm$, for $T_a$ from 4~K (lightest) to 300~K (darkest).}
\label{S11}
\end{figure}

Fig.~\ref{S11} demonstrates the measured input and output reflection coefficients $S_{11}$ and $S_{22}$ at $I_d = 25~\rm mA/mm$ and $V_{ds} = 0.5~\rm V$ which corresponded to the bias conditions used in the noise measurements (Sec. IV). For a given device, the $S_{11}$ plots for all $T_a$ almost overlap completely, and the three indium channel compositions closely trace similar paths. $S_{11}$ is essentially independent of $T_a$ and the channel indium content, as expected when it is dominated by the pad and access parasitics rather than by the intrinsic transistor. $S_{22}$ looks into the drain terminal and is dominated by the channel's output conductance $g_{ds}$. This parameter is set directly by the carrier transport in the channel which is strongly temperature-dependent as seen in $S_{22}$ in Fig.~\ref{S11}. 

\begin{figure}[!t]
\centering
\includegraphics[width=\columnwidth]{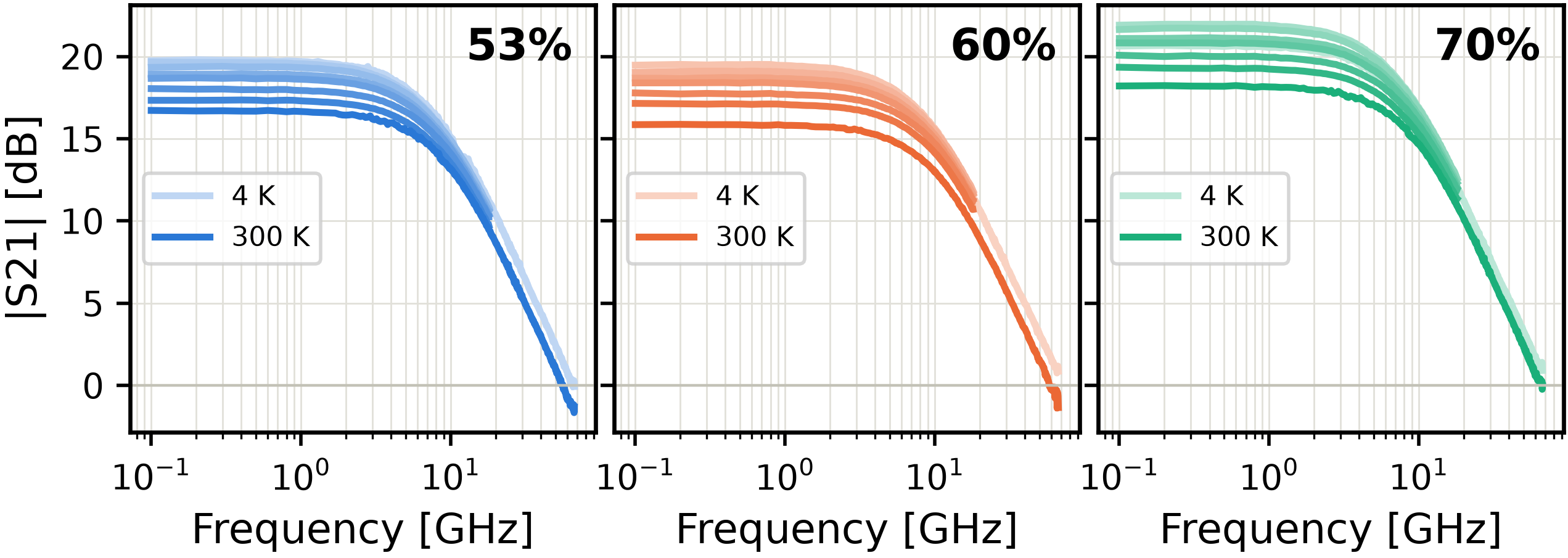}
\caption{Measured forward transmission $|S_{21}|$ versus frequency for 53\%, 60\% and 70\% indium channel InP HEMTs at $V_{ds} = 0.5~\rm V$ and
$I_d = 25~\rm mA/mm$. Trace shading indicates the ambient temperature from
$T_a = 4$~K (lightest) to 300~K (darkest).}
\label{S21}
\end{figure}

The forward transmission $|S_{21}|$ (same bias as Fig.~\ref{S11}) is shown in Fig.~\ref{S21}. All devices show a single-pole roll-off from 17-23~dB at 100~MHz to unity gain at 50-60~GHz. Cooling raises $|S_{21}|$ by 1-3~dB relative to 300~K, consistent with the increase in electron mobility and transconductance on cooling. The $|S_{21}|$ exhibits a strong dependence on indium channel: the 70\% indium channel InP HEMT shows the highest gain for all $T_a$, consistent with the approximately 1.5 dB higher gain observed for this device in the 4–8 GHz noise band reported in \cite{Li2024}.

Taken together, cooling these InP HEMTs produces a modest and consistent increase in gain and drain current without altering the input impedance environment presented. Any difference in noise performance between the three devices is therefore expected to originate from the intrinsic $\mathrm{In_{x}Ga_{1-x}As}$ channel rather than in the passive access or matching conditions.

\section{Noise characterization and discussion}
InP HEMTs were biased at $V_{ds}$ = 0.5 V and $I_d = 25~\rm mA/mm$. On-wafer, this bias was applied directly to the individual InP HEMT. In the LNA, the three stages were dc-biased in parallel from a common supply at $V_{DS} = 0.725~\rm V$ and $I_{D} = 15~\rm mA$. The difference between $V_{DS}$ and $V_{ds}$ is calculated from the bias network, so that each stage operated at $V_{ds} = 0.5~\rm V$ and $I_d = 25~\rm mA/mm$. The two noise measurement methods therefore probe the InP HEMT at an identical device bias point. This corresponded to the lowest noise at 4 K and provided sufficient gain also at 300 K\cite{Li2024,Ruiz2019}. The $V_{gs}$ was adjusted at each temperature to maintain constant bias conditions. While the InP HEMTs measured using the LNA had gate widths of 4×50 $\mu$m, the InP HEMTs probed on-wafer were 2×100 $\mu$m. The resulting difference in gate resistance caused by the different layouts had a small effect on the gate thermal noise which was not significant for estimation of the channel noise \cite{Cha2023}. The LNAs measured at 4 K had an average gain of 46 dB for 70\% indium channel, and 42 dB for 53\% and 60\% indium channel InP HEMTs. The average gain of the InP HEMTs measured on-wafer was between 15.0 and 17.5 dB for 80-300 K, with the 70\% indium channel exhibiting about 1.5 dB higher gain than 53\% and 60\% indium channel InP HEMTs. The amount of gain was sufficient for accurate noise estimation \cite{Ruiz2019}. 

For both methods of noise determination, the channel noise in the InP HEMT was extracted using the Pospieszalski model \cite{Marian1989}. The intrinsic output conductance $g_{ds,i}$ was obtained from the InP HEMT small-signal model using S-parameter measurements for each temperature and bias point \cite{Rorsman1996}. This procedure allowed extraction of the equivalent drain noise temperature $T_{d}$ which was used to calculate the channel current noise power spectral density (PSD) $S_{id} = 4k_{B}T_{d}g_{ds,i}$, where $k_{B}$ is the Boltzmann constant. For the LNA-based extraction, the package, bond wires, passive matching network were all fixed, and the input reference plane was set by a cold attenuator \cite{Cano2010}, leaving $T_d$ of the $\text{In}_x\text{Ga}_{1-x}\text{As}$ channel as the only free parameter. Because the three LNAs were identical in package, bond wires and matching network, their contributions to $T_{e,avg}$ were common to all three LNAs and bias-independent. As a result, $g_{ds,i}$ for the $\text{In}_x\text{Ga}_{1-x}\text{As}$ HEMT channel will be the dominant noise source at the optimum low-noise bias \cite{Li2024}. For $T_{a} \le $ 10 K, the gate noise temperature $T_{g}$ in Pospieszalski’s noise model was set to 10 K thus taking the device self-heating into account \cite{Ardizzi2022}. Above 10 K, $T_{g}$ was set equal to $T_{a}$. 

\begin{figure}[!t]
\centerline{\includegraphics[width=1\columnwidth]{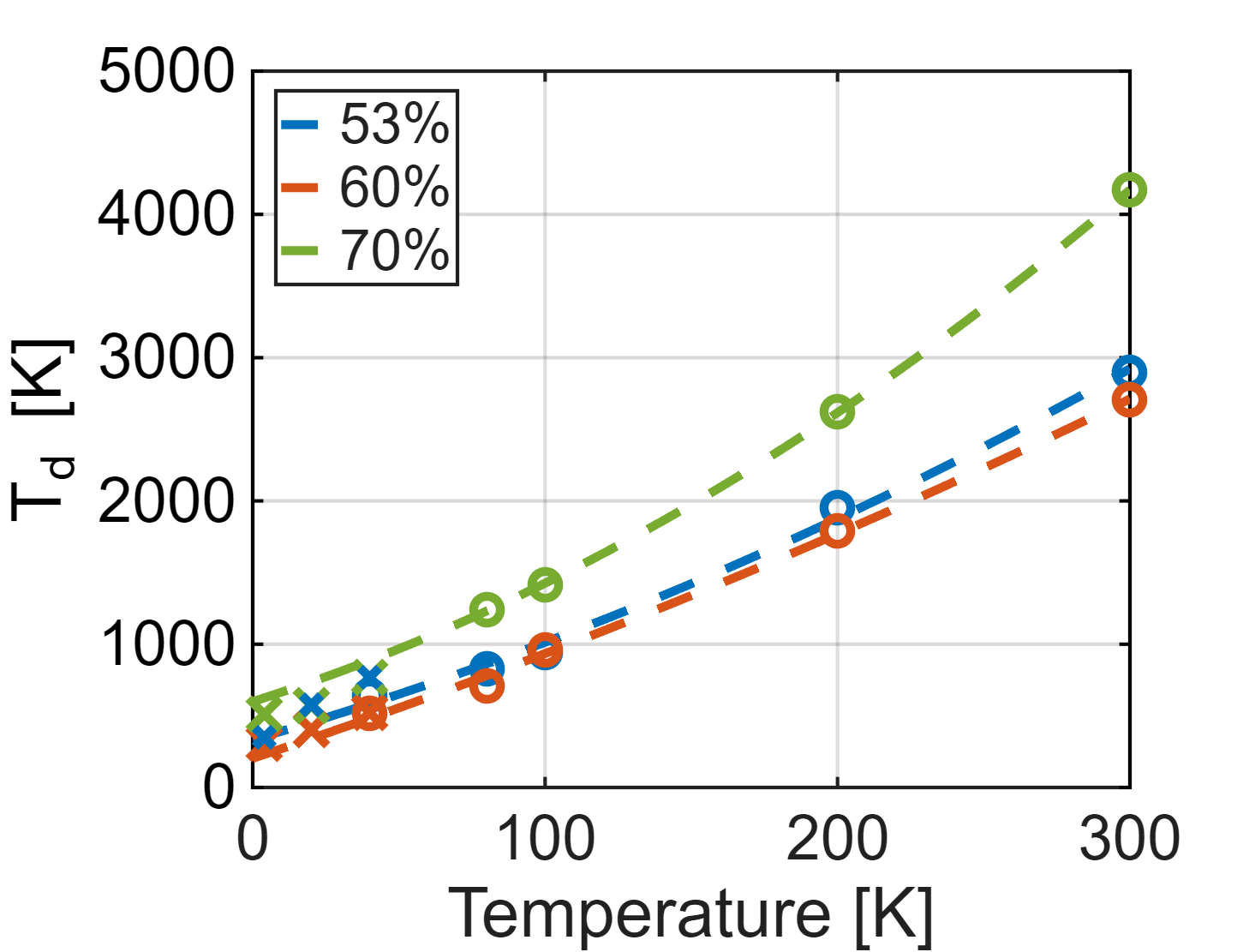}}
\caption{Equivalent drain noise temperature $T_{d}$ versus ambient temperature for InP HEMTs with channel indium content of 53\%, 60\%, and 70\%, biased at $V_{ds}$ = 0.5 V and $I_d = 25~\rm mA/mm$. Circles and crosses represent $T_{d}$ values extracted from on-wafer and LNA noise measurements, respectively. Dashed lines are parabolic fits to the on-wafer data above 40 K, extrapolated to 4 K. The absence of on-wafer data below 80 K for the 70\% indium channel InP HEMT was due to oscillations occurring during probing.}
\label{f1_1}
\end{figure}

The extracted $T_{d}$ from 4 K to 300 K is shown for all three InP HEMTs in Fig. \ref{f1_1}. Both measurement methods display consistent trends with good agreement between the on-wafer and LNA results for the 53\% and 60\% indium devices below 40 K. Across the full temperature range, $T_{d}$ is the lowest for the 60\% indium channel HEMT, confirming the indium content dependence with the $T_{e,avg}$ of the 4-8 GHz LNA at 4 K previously reported \cite{Li2024}. The observed increase of $T_{d}$ with $T_{a}$ is consistent with earlier studies \cite{Beka2023,Murti2000,Sandy2021,McCulloch2017,Ohmori2023,Schleeh2013}. Even though $T_{d}$ also has been reported to be weakly dependent on temperature \cite{Pospieszalski2017}, it appears that for the low $I_{d}$ applied here, a parabolic fit to the on-wafer data above 40 K provides a good description of the temperature dependence \cite{Munoz_1997}; See dashed lines in Fig. \ref{f1_1}. Furthermore, the tendency for the saturation of $T_{d}$ at low temperatures is probably influenced by device self-heating \cite{Schleeh2015}. Higher dc power dissipation is expected to result in stronger $T_{d}$ saturation as the $T_{a}$ decreases, consistent with the weaker saturation observed for $T_{d}$ in our devices compared to those in \cite{Munoz_1997}. 

\begin{figure}[!t]
\centerline{\includegraphics[width=1\columnwidth]{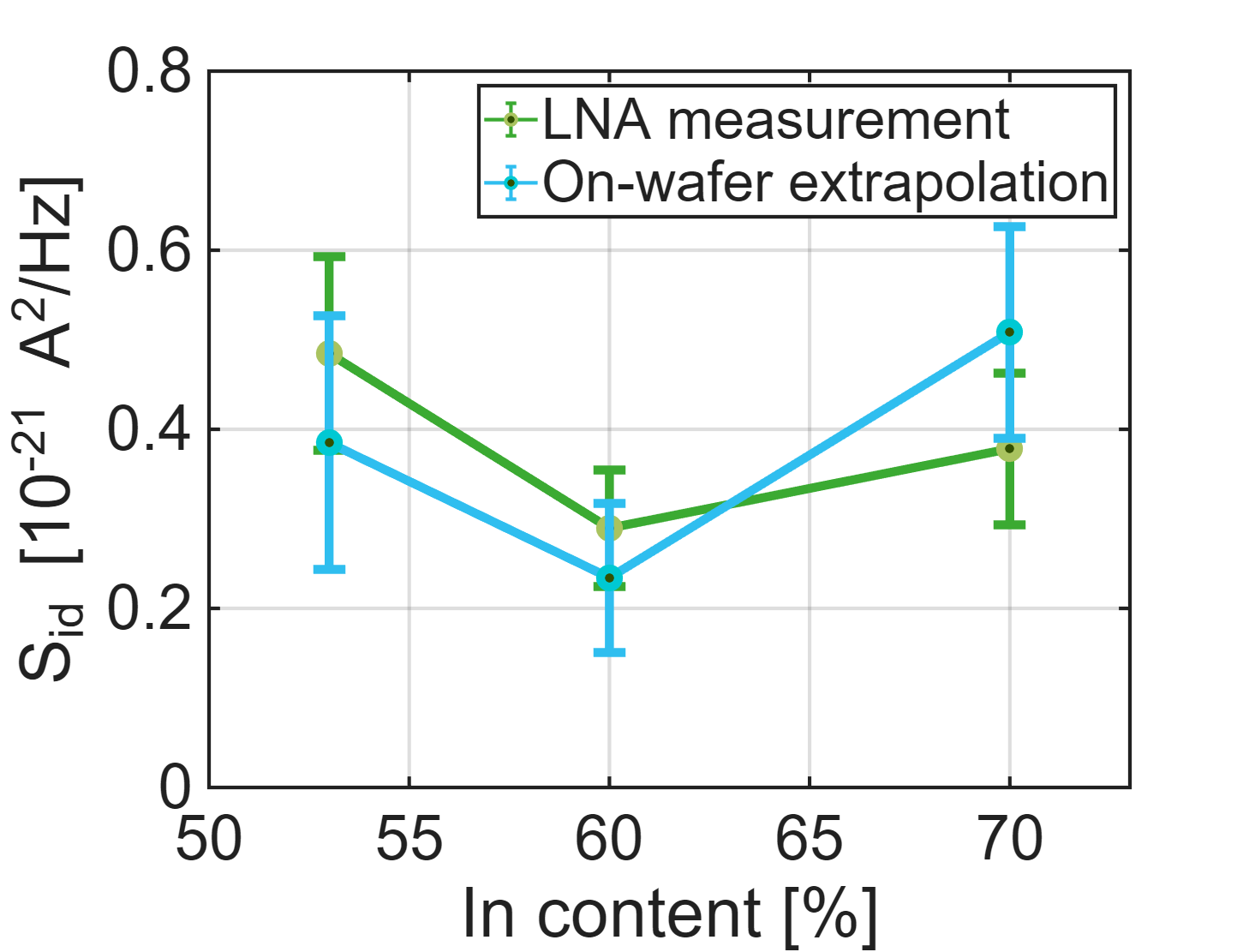}}
\caption{ Channel current noise spectral density $S_{id}$ at 4 K as a function of indium content x in $\mathrm{In_{x}Ga_{1-x}As}$ channel InP HEMTs. Green points show values extracted from LNA noise measurements. Blue points show values extrapolated to 4 K from the $T_{d}$ model fitted to on-wafer measurements (Fig. 3); Error bars represent the combined measurement uncertainties and fitting errors.}
\label{f2_1}
\end{figure}

The $S_{id}$ at 4 K versus channel indium content is shown in Fig. \ref{f2_1}, comparing values obtained from the LNA measurements and the parabolic extrapolation of the on-wafer noise data to 4 K in Fig. \ref{f1_1}. Both noise measurement methods display the same trend in $S_{id}$ for varying channel indium content in the InP HEMT. This demonstrates that cryogenic noise performance at 4 K can be predicted without the need for dedicated cryogenic characterization at this temperature. It is confirmed that the 60\% indium channel InP HEMT exhibits the lowest channel noise at cryogenic temperatures. 

The observed temperature dependence of $T_{d}$ in Fig. \ref{f1_1} indicates that the channel noise cannot be attributed only to suppressed shot noise \cite{Marian2017}. The temperature dependence of $T_{d}$ in a suppressed shot noise  model can be expressed by:
\begin{equation}
    T_d = \frac{qF I_{d}}{2k_{B} g_{ds,i}},
\end{equation}
where $F$ is the Fano factor \cite{Das_2025}. Assuming $F$ is temperature independent, Eq. (1) predicts a two-fold change in $T_{d}$ due to the variation of $g_{ds,i}$ between 4 K and 300 K for the InP HEMTs studied here. However, Fig. \ref{f1_1} demonstrates an increase with a factor of 6 in $T_{d}$ from 4 K to 300 K.  Similar discrepancies have been reported in \cite{Heinz_2022} and also seen in the data in \cite{Schleeh2013}. Hence the noise measurements in this study strongly suggest that the temperature dependence of $T_{d}$ in the InP HEMTs reflects a physical mechanism for the channel noise beyond suppressed shot noise.

\begin{figure}[!t]
\centerline{\includegraphics[width=1\columnwidth]{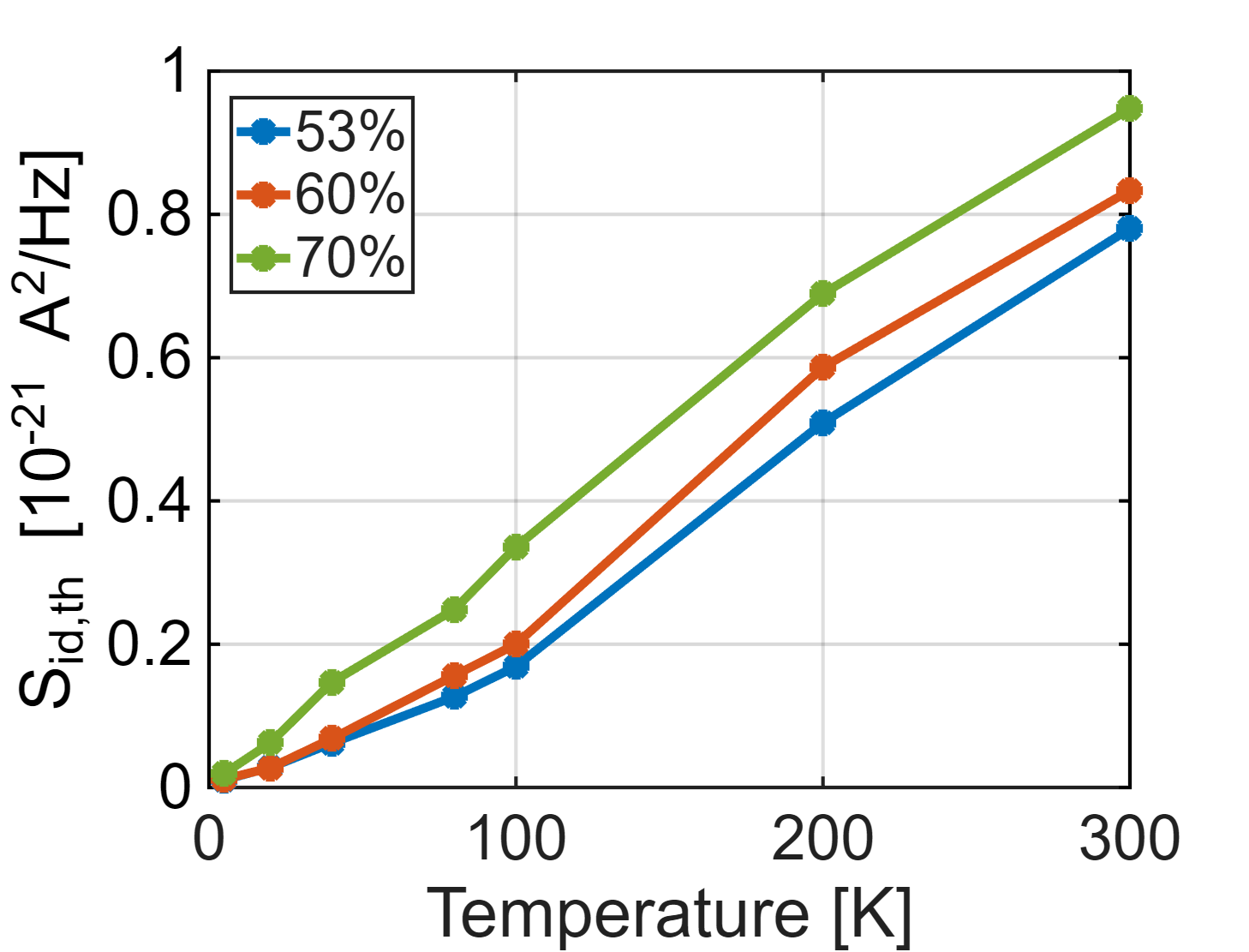}}
\caption{Thermal noise PSD $S_{id,th} = 4k_{B}T_{a}g_{ds0}$ versus ambient temperature for InP HEMTs with channel indium content of 53\%, 60\%, and 70\%.}
\label{f2_2}
\end{figure}

\begin{figure}[!t]
\centerline{\includegraphics[width=1\columnwidth]{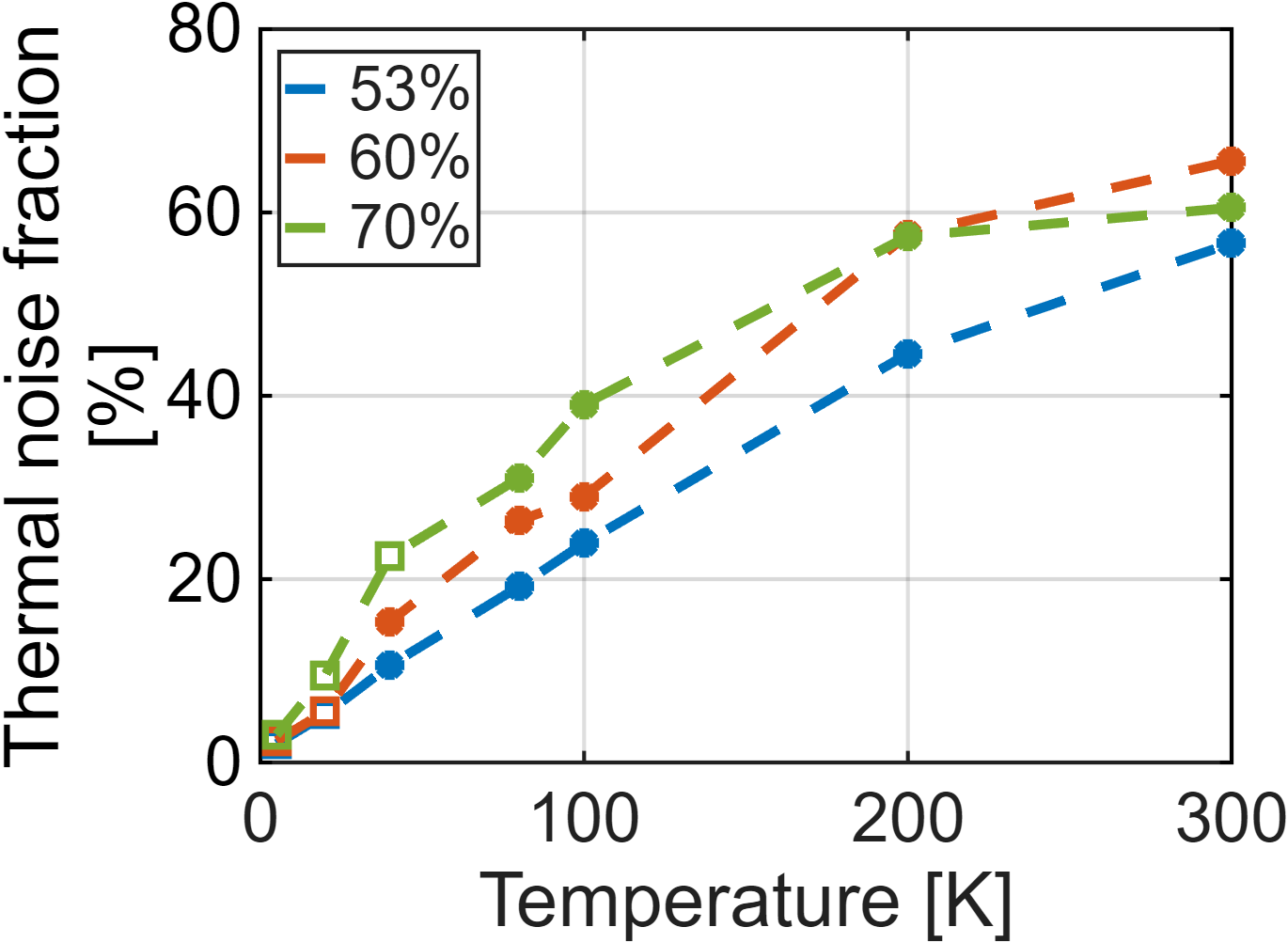}}
\caption{Fractional contribution of thermal noise to the total channel noise PSD $S_{id,th}/S_{id}$ versus ambient temperature for InP HEMTs with channel indium content of 53\%, 60\%, and 70\%.}
\label{f2_3}
\end{figure}

To analyze the origin of the channel noise, $S_{id}$ can be decomposed into a thermal noise contribution $S_{id,th}$ and an excess noise contribution $S_{id,ex}$ \cite{Beka2023}. The thermal noise PSD is expressed by $S_{id,th} = 4k_{B}T_{a}g_{ds0}$, where $g_{ds0}$ is the output conductance at zero drain bias extracted from the $I_{d}-V_{ds}$ characteristics \cite{Ziel1962}. $S_{id,th}$ represents the equilibrium thermal noise contribution, corresponding to channel electrons in thermal equilibrium with the lattice at $T_{a}$. In 100-nm gate-length InP HEMTs operating in saturation, carriers near the drain end of the channel are heated significantly above $T_{a}$, with an electron temperature $T_e$ typically above 1000 K \cite{Hartnagel2001, Tomi2022}. The channel noise originating from these hot electrons is captured in $S_{id,ex}$. The $S_{id,th}$ provides a lower bound for the thermal noise. $S_{id,th}$ is shown versus $T_{a}$ in Fig. \ref{f2_2} and reveals a linear temperature dependence. $S_{id,th}$ increases with indium content, as higher indium content yields higher electron mobility and thus higher $g_{ds0}$ \cite{Li2024}. The fractional contribution of $S_{id,th}$ to the total channel noise $S_{id}$, illustrated in Fig. \ref{f2_3}, confirms that the thermal noise accounts for the majority of $S_{id}$ at room temperature and becomes negligible at 4 K.

\begin{figure}[!t]
\centerline{\includegraphics[width=1\columnwidth]{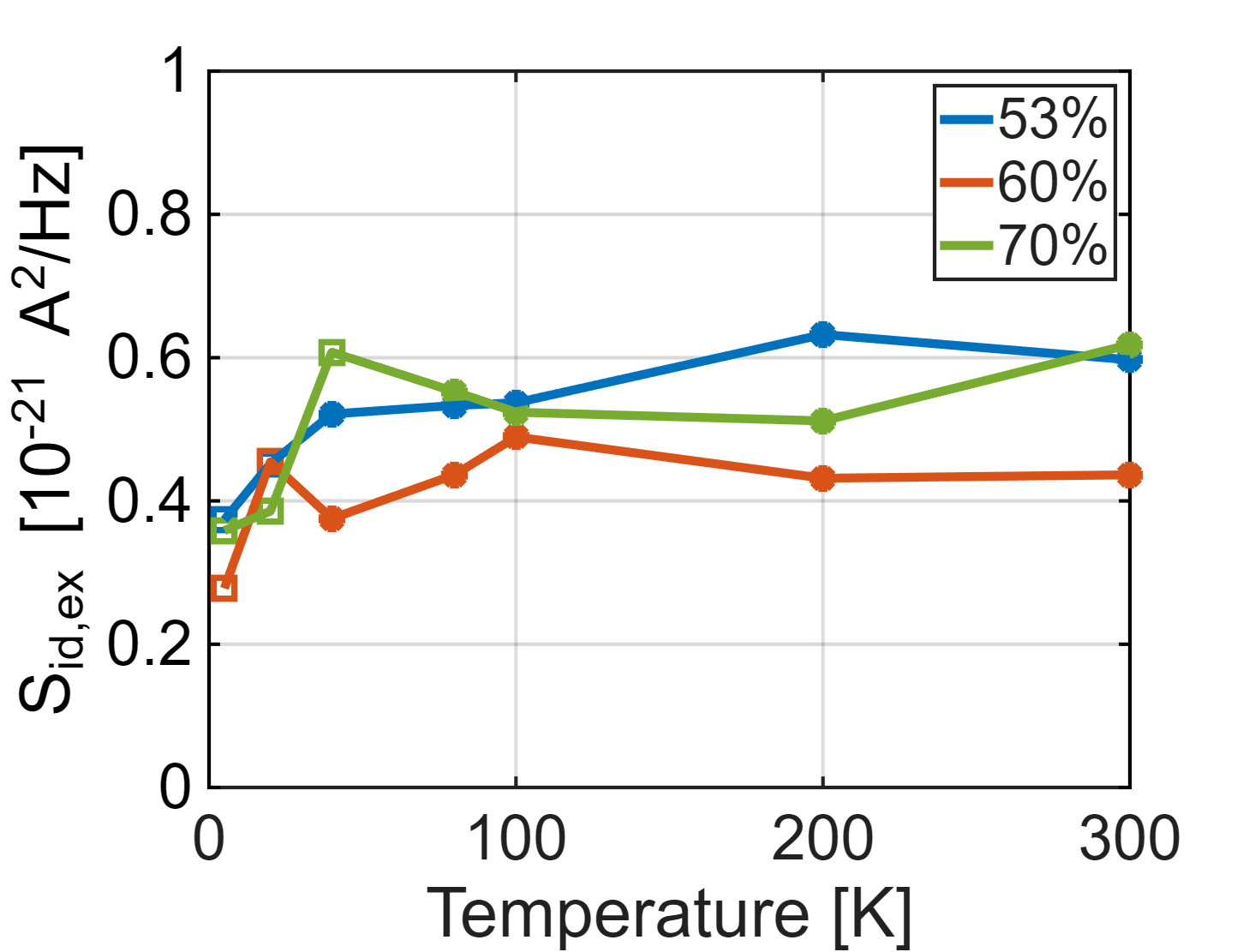}}
\caption{Excess noise PSD $S_{id,ex}$ versus ambient temperature for InP HEMTs with channel indium content of 53\%, 60\%, and 70\%. Solid markers: values extracted from on-wafer noise measurements; open markers: values extracted from LNA measurements.}
\label{f3_1}
\end{figure}

The excess noise $S_{id,ex} =  S_{id} - S_{id,th}$, plotted versus $T_{a}$ in Fig. \ref{f3_1}, remains essentially independent of temperature for all three indium compositions, confirming that $S_{id,ex}$ dominates the total channel noise at 4 K \cite{LiCSW2024}. The temperature independence of $S_{id,ex}$ is consistent with a shot-noise-like origin. The Fano factor for excess noise $F_{ex}=\frac{\overline{S_{id,ex}}}{2qI_d}$, where $\overline{S_{id,ex}}$ denotes averaging over the measured temperature range 4-300 K, exhibits values of 0.33, 0.26, and 0.32 for the 53\%, 60\%, and 70\% indium channels, respectively. The 60\% indium channel shows the lowest excess noise, with $F_{ex}$ approximately 20\% lower than the other two channel compositions. In the suppressed shot noise model, $F$ is determined by the gate length and $I_{d}$ \cite{Marian2017, Das_2025}. Since gate length and $I_{d}$ are identical for all three devices, the observed variation of $S_{id,ex}$ with indium content can not be exclusively attributed to a suppressed shot noise mechanism. 

We now discuss two hot-electron mechanisms which may be involved in generation of channel noise for the InP HEMTs: impact ionization, in which energetic channel carriers generate additional electron-hole pairs \cite{Reuter1997}, and RST where hot electrons are emitted from the $\mathrm{In_{x}Ga_{1-x}As}$ channel into the InAlAs barrier \cite{Hartnagel2001}. Both noise mechanisms depend on the channel indium composition of the InP HEMT but in different ways.

\begin{figure}[!t]
\centerline{\includegraphics[width=1\columnwidth]{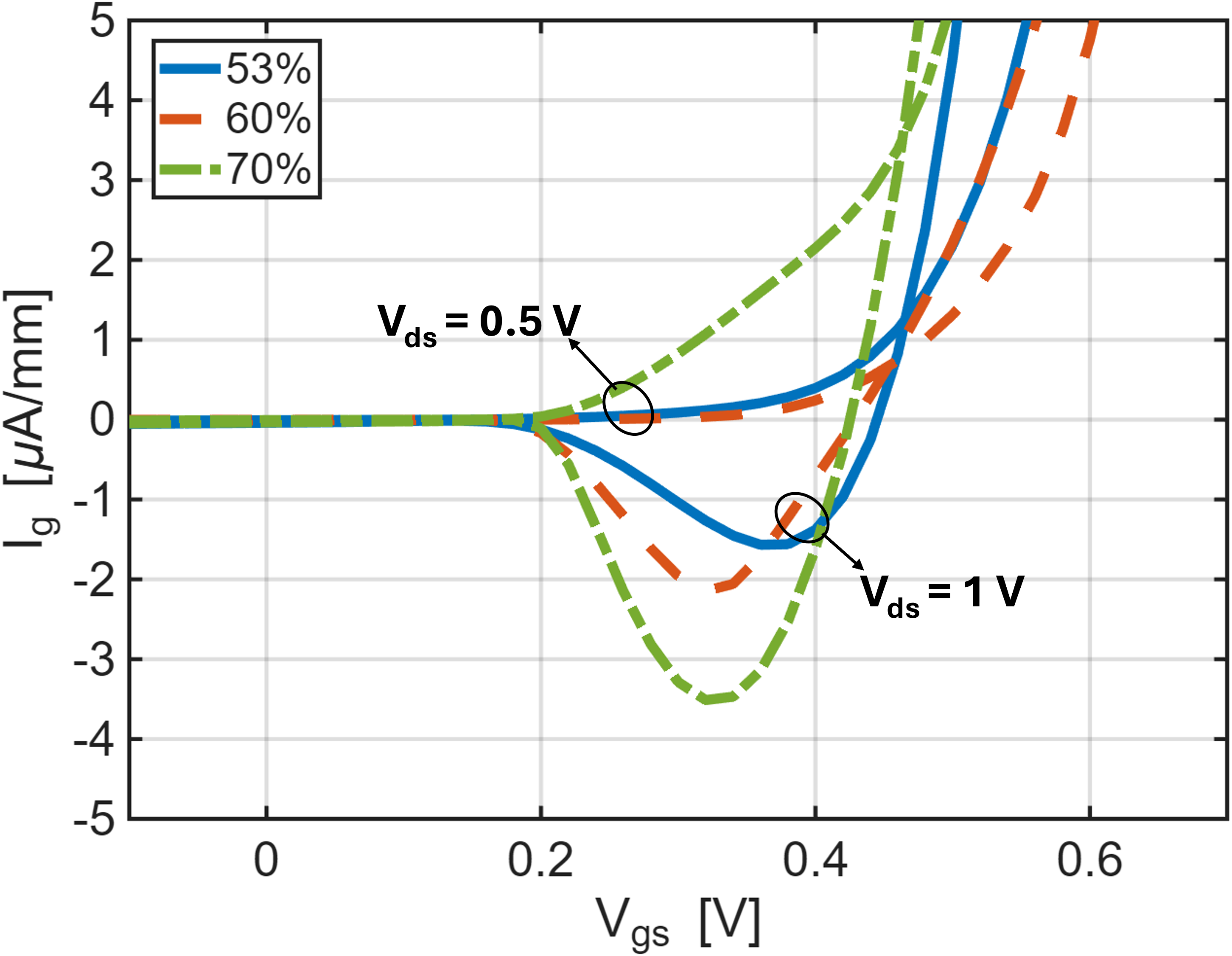}}
\caption{Gate current $I_{g}$ versus $V_{gs}$ at $V_{ds}$ = 0.5 V and 1 V at 4 K for InP HEMTs with channel indium content of 53\%, 60\%, and 70\%.}
\label{f3_3}
\end{figure}

\begin{figure}[!t]
\centerline{\includegraphics[width=1\columnwidth]{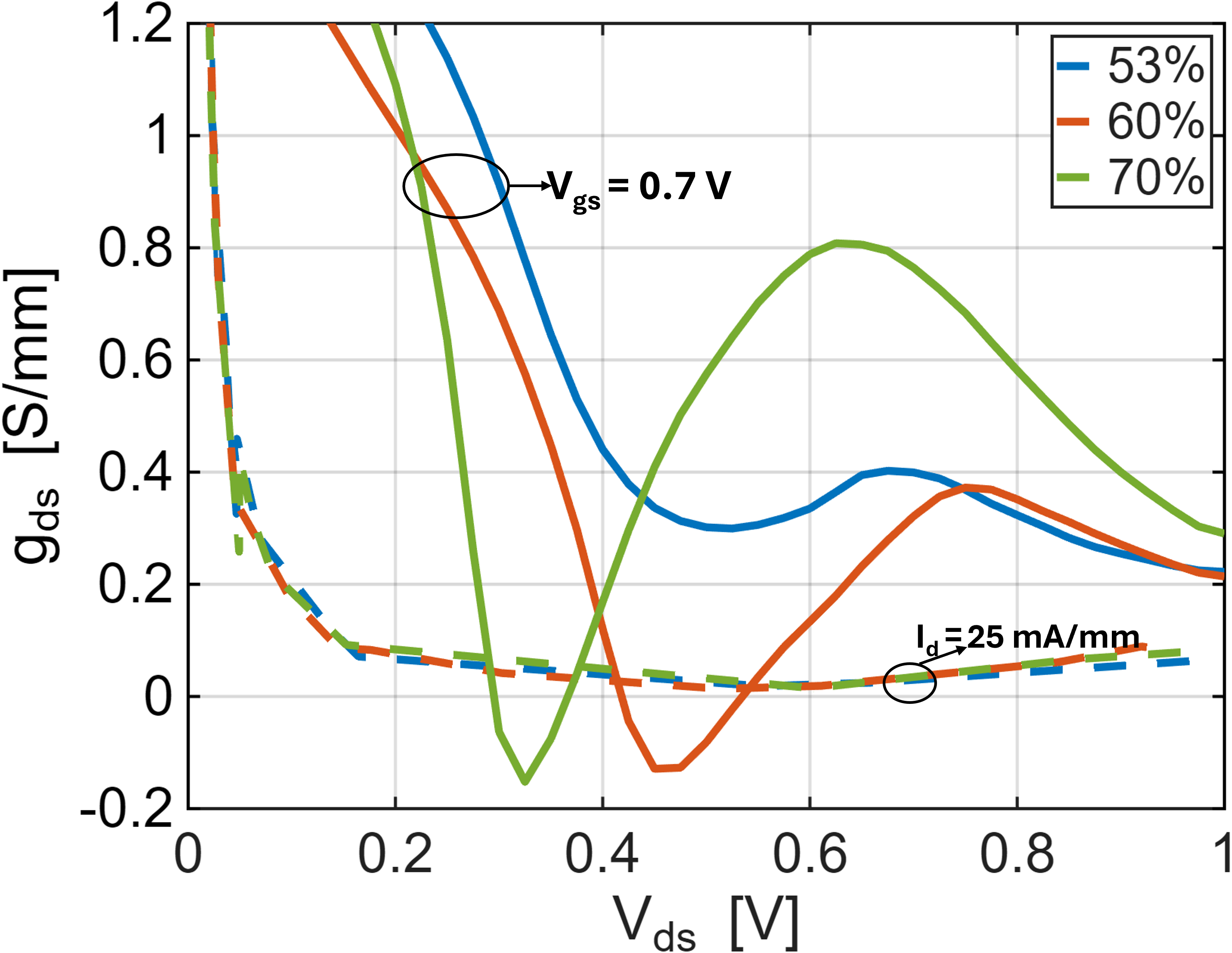}}
\caption{Output conductance $g_{ds}$ versus $V_{ds}$ at 4~K for the 53\%, 60\% and 70\% indium channel InP HEMTs. Dashed lines: $V_{gs}$ adjusted at each $V_{ds}$ to keep $I_d = 25~\rm mA/mm$, the noise-measurement bias. Solid lines: fixed $V_{gs} = 0.7~\rm V$.}
\label{f3_2}
\end{figure}

Impact ionization requires the charge carriers in the 2DEG to acquire sufficient energy to generate electron-hole pairs. In 100-nm gate-length InP HEMTs, Monte Carlo simulations indicates that the peak electric field under the gate exceeds this energy even at the low-noise bias point for 53\% indium channel \cite{Mateos1999}. Impact ionization is strongly dependent on the bandgap $E_{g}$ of the $\mathrm{In_{x}Ga_{1-x}As}$ channel \cite{Pearsall1978}. The $E_{g}$ at 300 K of the channel follows the bowing parameterization\cite{Vurgaftman2001}:
\begin{equation}
E_g(x) = (1 - x)E_{g,\text{GaAs}} + x E_{g,\text{InAs}} - C x(1 - x),
\quad
\label{eq:bandgap}
\end{equation}
with $C = 0.477 \text{ eV}$, $E_{g,\text{GaAs}} = 1.52$ eV and $E_{g,\text{InAs}} = 0.42$ eV \cite{Vurgaftman2001}, Eq.~\eqref{eq:bandgap} gives $E_g = 0.82$, $0.74$ and $0.65$ eV for $x = 0.53$, $0.60$ and $0.70$, respectively. Since the 60\% and 70\% channels are pseudomorphically strained, these values are a conservative lower bound on $E_g$. For the InP HEMTs studied here, the 70\% indium channel device has the narrowest $E_{g}$ and will suffer from the strongest impact ionization noise \cite{Pearsall1978}. This is confirmed by the measured gate current $I_{g}$ at the low-noise bias ($V_{ds}$ = 0.5 V, $V_{gs} = 0.22-0.26$ V for $I_{d}$ = 25 mA/mm) plotted in Fig. \ref{f3_3}, where the 70\% device shows one order of magnitude higher $I_{g}$ (80 nA at 4 K, 110 nA at 300 K) than the 53\% and 60\% devices (5 and 3 nA at 4 K, 10 and 30 nA at 300 K, respectively). Although $I_{g}$ itself is too small to significantly contribute to the channel noise via gate shot noise ($2qI_{g} \sim 10^{-26} \mathrm{A^2/Hz}$), its elevated value is a well-established indicator of enhanced impact ionization noise \cite{Gaquiere1999}. The bell-shaped curves for $I_{g}$ at  $V_{ds} = $ 1 V is also a characteristic for impact ionization \cite{Zhou2024} accompanied by an increase in $g_{ds}$ extracted from the $I_{d}-V_{ds}$ characteristics as shown in Fig. \ref{f3_2}. The presence of impact ionization taking place in the InP HEMT may create excess noise in the channel also under the low-noise bias conditions used here as suggested by Monte Carlo simulations \cite{Vasallo2004} and experimental data \cite{RuizIEDM}. The elevated excess noise observed for the 70\% indium channel device may therefore be associated with a stronger noise contribution originating from impact ionization taking place in the channel of the InP HEMT.

\begin{figure}[t]
\centerline{\includegraphics[width=1\columnwidth]{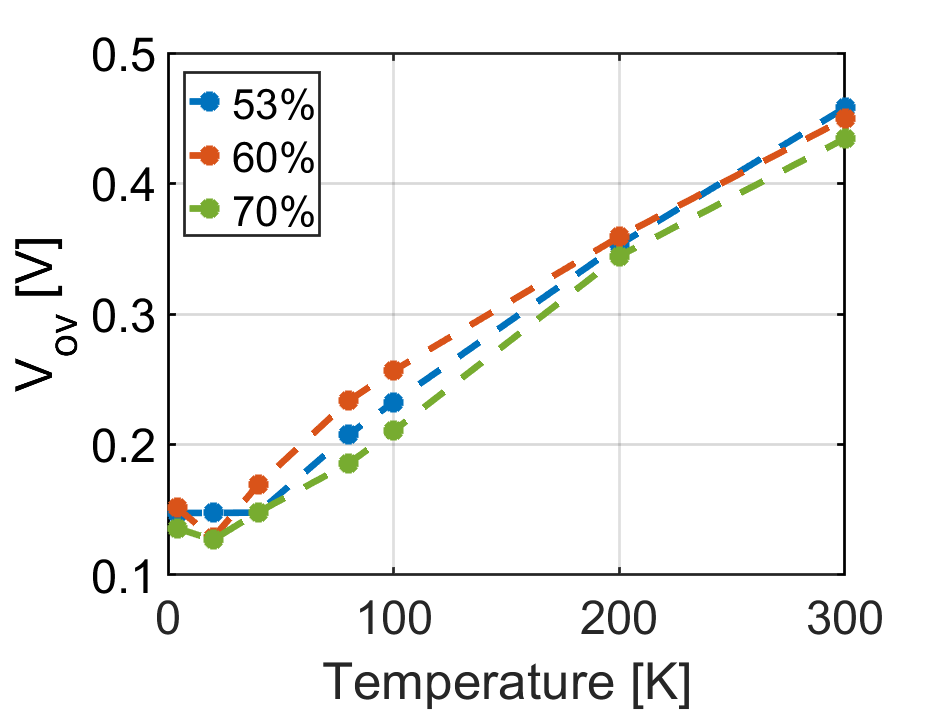}}
\caption{Overdrive voltage $V_{ov}$ versus ambient temperature for InP HEMTs with channel indium content of 53\%, 60\%, and 70\%.}
\label{Vov}
\end{figure}

RST noise in the InP HEMT arises because the electrical transport properties differ substantially between the $\mathrm{In_{x}Ga_{1-x}As}$ channel and the InAlAs barrier \cite{Hartnagel2001}. Following \cite{Tomi2022}, the RST contribution to the drain noise temperature can be written as
\begin{equation}
T_{d,\text{RST}} = \frac{q^2 v_d^2 n_{s,2} \tau_{\text{RST}}}{k_B g'_{ds,i} L},
\quad
n_{s,2} = \gamma \eta n_{s,1},
\quad
\label{eq:rst_noise}
\end{equation}
where $v_d$ is the electron drift velocity in the channel, $\tau_{\text{RST}}$ the characteristic channel-to-barrier transfer time, $g'_{ds,i} = g_{ds,i}/W$ the intrinsic output conductance per unit gate width, $L$ the length of the hot-electron region at the drain end of the channel, $n_{s,1}$ the channel sheet carrier density in that region, $n_{s,2}$ the sheet density of electrons transferred into the barrier, $\gamma$ the emission probability. $\eta$ is the fraction of electrons that transfer from the channel to the barrier depends exponentially on the conduction band offset $\Delta E_{c}$ at the InGaAs/InAlAs interface through $\eta = \exp\left[-\frac{\Delta E_c - q V_{ov}}{k_B T_e}\right]$. In \cite{Tomi2022} the exponent is written with $V_{gs}$; here it is expressed through the overdrive voltage $V_{ov} = V_{gs} - V_{th}$, where $V_{th}$ is the threshold voltage, thus allowing fair comparison between different devices \cite{Li2022}. Since $V_{ov}$ is similar for the three devices (Fig. \ref{Vov}), $V_{ov}$ contribution to the RST noise largely cancels in a comparison between them. Evaluating $\exp(-\Delta E_c/k_B T_e)$ at $T_e = 1000$ K \cite{Tomi2022, Hartnagel2001} using the $\Delta E_c$ values above yields $2.4 \times 10^{-3}$, $6.0 \times 10^{-4}$ and $3.7 \times 10^{-4}$ for the 53\%, 60\% and 70\% indium channels, respectively \cite{Huang1992}. This is a four-fold reduction of the transferred fraction from 53\% to 60\% indium and a further 1.6-fold reduction from 60\% to 70\%. The remaining quantities in Eq.~\eqref{eq:rst_noise} vary much more weakly with channel indium content than this exponential, meaning that $\Delta E_c$ governs the composition dependence of the RST noise. Indeed, recent experimental evidence by varying barrier compositions in InGaAs/InAlAs quantum wells has confirmed that RST contributes to microwave noise, supporting the hypothesis that RST noise is part of the channel noise in our InP HEMTs \cite{Zhang2026}.

Both RST and impact ionization require hot electrons with $T_e \gg T_a$. Since $T_e$ is primarily determined by the electric field under low-noise bias\cite{Tomi2022, Mateos1999}, the variation of $T_a$ from 4 K to 300 K represents a negligible perturbation to $T_e$ compared to the field-induced heating. Both Eqs.~\eqref{eq:bandgap} and ~\eqref{eq:rst_noise} depend on the channel indium content with opposite sign. As a result, the non-monotonic dependence of excess noise on channel indium content in Fig. \ref{f3_1} can be understood from the combination of RST and impact ionization both contributing to channel noise. For the 53\% indium channel device, the small $\Delta E_{c}$ means higher probability for RST, elevating the excess noise. For the 70\% indium channel device, the narrow bandgap promotes impact ionization, also increasing the excess noise. The 60\% indium channel device achieves the optimal balance between impact ionization and RST noise resulting in the lowest $S_{id,ex}$.

\section{Conclusion}
We have investigated the channel noise from 4 K to 300 K for 100-nm gate-length InP HEMTs with $\mathrm{In_{x}Ga_{1-x}As}$ channel indium content ranging from 53\% to 70\%. Two different noise measurement methods were utilized, one based on on-wafer probing above 40 K and one using an LNA below 40 K. The two methods yielded consistent results across the full temperature range, confirming the reliability of the noise characterization. The 60\% indium channel InP HEMT exhibited the lowest $T_d$  across the entire temperature range. The on-wafer noise data above 40 K was fitted with a parabolic function and extrapolated to 4 K, showing good agreement with the LNA measurement results. The channel noise comprised a thermal noise contribution dominating at 300 K and a temperature-independent excess noise contribution dominating at 4 K. The non-monotonic dependence of $S_{id,ex}$ on channel indium content for fixed gate length and bias could not be explained by only taking suppressed shot noise into account. It is therefore suggested that excess noise in the channel also must be attributed to additional hot-electron mechanisms, specifically impact ionization and RST. These noise mechanisms exhibit opposite dependencies on channel indium content through $\Delta E_c$ and $E_g$, resulting in the observed minimum excess noise for the 60\% indium channel HEMTs. The results suggest that further advancement of InP HEMT LNA performance for radio astronomy and quantum computing should focus on reducing channel noise not only from suppressed shot noise but also from impact ionization and RST noise.

\section*{Acknowledgment}

The authors extend their gratitude to Estefany Santana, Johan Bergsten, Arsalan Pourkabirian, Jörgen Stenarson and Niklas Wadefalk at Low Noise Factory AB. Additionally, the authors are thankful to Bekari Gabritchidze, Jacob Kooi at JPL, Kieran Cleary, Anthony J. Ardizzi, Jiayin Zhang at Caltech and Marian W. Pospieszalski at NRAO for engaging in fruitful discussions. This work was supported by the WiTECH Centre project CRYTER+ through Vinnova, Low Noise Factory AB, Virgina Diodes Inc., AAC Omnisys, and RISE Research Institutes of Sweden.

\newpage

\vfill


\begin{thebibliography}{99}

\bibliographystyle{IEEEtran}

\bibitem{Bardin2021}
J. C. Bardin, ``Cryogenic Low-Noise Amplifiers: Noise Performance and Power Dissipation,'' \emph{IEEE Solid-State Circuits Mag.}, vol. 13, no. 2, pp. 22--35, Spring 2021, doi: 10.1109/MSSC.2021.3072803.

\bibitem{Bryerton2013}
E. W. Bryerton, M. Morgan, and M. W. Pospieszalski, ``Ultra low noise cryogenic amplifiers for radio astronomy,'' 
\emph{Proc. IEEE Radio Wireless Symp.}, Austin, TX, USA, 2013, pp. 358--360, doi: 10.1109/RWS.2013.6486740.

\bibitem{Hornibrook2015} 
J. M. Hornibrook, J. I. Colless, I. D. Conway Lamb, S. J. Pauka, H. Lu, A. C. Gossard, J. D. Watson, G. C. Gardner, S. Fallahi, M. J. Manfra, D. J. Reilly,
{“Cryogenic control architecture for large-scale quantum computing”},
\emph{Phys. Rev. Appl.},
vol. 3, no. 2, 2015,
doi:10.1103/PhysRevApplied.3.024010.

\bibitem{HelenaTED}
H. Rodilla, J. Schleeh, P.-Å. Nilsson, N. Wadefalk, J. Mateos, and J. Grahn, ``Cryogenic Performance of Low-Noise InP HEMTs: A Monte Carlo Study,'' \emph{IEEE Trans. Electron Devices}, vol. 60, no. 5, pp. 1625--1631, May 2013, doi: 10.1109/TED.2013.2253469.

\bibitem{Cha2023} 
E. Cha, N. Wadefalk, G. Moschetti, A. Pourkabirian, J. Stenarson, J. Li, D.-H. Kim, and J. Grahn,
``Optimization of channel structures in InP HEMT technology for cryogenic low-noise and low-power operation,''
\emph{IEEE Transactions on Electron Devices},
vol. 70, no. 5, pp. 2431-2436, 2023.
doi: 10.1109/TED.2023.3255160.

\bibitem{Schleeh2012}
J. Schleeh et al., ``Ultralow-Power Cryogenic InP HEMT With Minimum Noise Temperature of 1 K at 6 GHz,'' \emph{IEEE Electron Device Lett.}, vol. 33, no. 5, pp. 664--666, May 2012, doi: 10.1109/LED.2012.2187422.

\bibitem{Rebelo2026}
N. Rebelo, J. Bergsten, A. Pourkabirian, N. Wadefalk, and J. Grahn, ``A Low-Power Cryogenic Low-Noise Amplifier for the Next-Generation Quantum Computers,'' \emph{Phys. Status Solidi A}, vol. 223, pp. e202501018, 2026, doi: 10.1002/pssa.202501018.

\bibitem{Li2024} 
J. Li, J. Bergsten, A. Pourkabirian, and J. Grahn,
``Investigation of noise properties in the InP HEMT for LNAs in qubit amplification: effects from channel Indium content,''
\emph{IEEE Journal of the Electron Devices Society},
vol. 1, no. 1, pp. 1-1, 2024.
doi: 10.1109/JEDS.2024.3371905.


\bibitem{Cha2018} 
E.  Cha,  N.  Wadefalk,  P.  Nilsson,  J.  Schleeh,  G.  Moschetti,  A.  Pourkabirian,  S.  Tuzi,   and J. Grahn, 
{“0.3–14 and 16–28 GHz wide-bandwidth cryogenic MMIC low-noise amplifiers”},
\emph{IEEE Transactions on Microwave Theory and Techniques},
vol. 66, no. 11, pp.4860-4869, 2018,
doi:10.1109/TMTT.2018.2872566.

\bibitem{Bryerton2009}
E. W. Bryerton, X. Mei, Y.-M. Kim, W. Deal, W. Yoshida, M. Lange, J. Uyeda, M. Morgan, and R. Lai, ``A W-band low-noise amplifier with 22K noise temperature,'' in \emph{Proc. IEEE MTT-S Int. Microw. Symp.}, Boston, MA, USA, 2009, pp. 681--684, doi: 10.1109/MWSYM.2009.5165788.

\bibitem{Heinz2021}
F. Heinz, F. Thome, A. Leuther, and O. Ambacher, ``A 50-nm Gate-Length Metamorphic HEMT Technology Optimized for Cryogenic Ultra-Low-Noise Operation,'' \emph{IEEE Trans. Microw. Theory Techn.}, vol. 69, no. 8, pp. 3896--3907, Aug. 2021, doi: 10.1109/TMTT.2021.3081710.

\bibitem{Heinz2026}
F. Heinz, F. Thome, P. Pütz, G. Wieching, and A. Leuther, ``A 16--50 GHz Cryogenic LNA MMIC with 11.2 K Average Noise Temperature for Radio Astronomy,'' in \emph{Proc. IEEE MTT-S Radio Freq. Technol. Techn. Symp. (IMS RFTT)}, Boston, MA, USA, 2026, pp. 257--260, doi: 10.1109/IMSRFTT43070.2026.11602539.

\bibitem{ChaIMS} 
E. Cha, N. Wadefalk, G. Moschetti, A. Pourkabirian, J. Stenarson, and J. Grahn,
``A 300-$\mu$W cryogenic {HEMT} {LNA} for quantum computing,''
in \emph{2020 IEEE/MTT-S International Microwave Symposium (IMS)},
2020, pp. 1299-1302.
doi: 10.1109/IMS30576.2020.9223865.


\bibitem{Ruiz2019} 
D. C. Ruiz, T. Saranovac, D. Han, A. Hambitzer, A. M. Arabhavi, O. Ostinelli, and C. R. Bolognesi,
``InAs channel inset effects on the DC, RF, and noise properties of InP pHEMTs,''
\emph{IEEE Transactions on Electron Devices},
vol. 66, no. 11, pp. 4685-4691, 2019.
doi: 10.1109/TED.2019.2940638.

\bibitem{Wang2022}
C. Wang, C.-N. Kuo, Y.-C. Lin, H.-T. Hsu, Y.-F. Tsao, C.-T. Lee, and E. Y. Chang, ``Effect of the Indium Compositions in Tri-Gate In$_x$Ga$_{1-x}$As HEMTs for High-Frequency Low Noise Application,'' \emph{ECS J. Solid State Sci. Technol.}, vol. 11, no. 11, pp. 115006, Nov. 2022, doi: 10.1149/2162-8777/aca04d.


\bibitem{Smit2014}
G. D. J. Smit, A. J. Scholten, R. M. T. Pijper, L. F. Tiemeijer, R. van der Toorn, and D. B. M. Klaassen, ``RF-Noise Modeling in Advanced CMOS Technologies,'' \emph{IEEE Trans. Electron Devices}, vol. 61, no. 2, pp. 245--254, Feb. 2014, doi: 10.1109/TED.2013.2282960.

\bibitem{Marian2005}
M. W. Pospieszalski, ``Extremely low-noise amplification with cryogenic FETs and HFETs: 1970-2004,'' \emph{IEEE Microw. Mag.}, vol. 6, no. 3, pp. 62--75, Sep. 2005, doi: 10.1109/MMW.2005.1511915.


\bibitem{Marian2017} 
M. W. Pospieszalski,
``On the limits of noise performance of field effect transistors,''
in \emph{2017 IEEE MTT-S International Microwave Symposium (IMS)},
2017, pp. 1953-1956.
doi: 10.1109/MWSYM.2017.8059045.

\bibitem{Tomi2022} 
I. Esho, A. Y. Choi, and A. J. Minnich,
``Theory of drain noise in high electron mobility transistors based on real-space transfer,''
\emph{Journal of Applied Physics},
vol. 131, no. 8, article number: 085111, 2022.
doi: 10.1063/5.0069352.

\bibitem{Li2022} 
J. Li, A. Pourkabirian, J. Bergsten, N. Wadefalk, and J. Grahn,
``Influence of spacer thickness on the noise performance in InP HEMTs for cryogenic LNAs,''
\emph{IEEE Electron Device Letters},
vol. 43, no. 7, pp. 1029-1032, 2022.
doi: 10.1109/LED.2022.3178613.


\bibitem{RuizIEDM} 
D. C. Ruiz, T. Saranovac, D. Han, O. Ostinelli, C.R. Bolognesi,
{“Impact ionization control in 50 nm low-noise high-speed {InP} {HEMTs} with {InAs} channel insets”},
\emph{2019 IEEE International Electron Devices Meeting (IEDM)}, 2019,
doi:10.1109/IEDM19573.2019.8993654.


\bibitem{Wadefalk2003}
N. Wadefalk, A. Mellberg, I. Angelov, M. E. Barsky, S. Bui, E. Choumas, R. W. Grundbacher, E. L. Kollberg, R. Lai, N. Rorsman, P. Starski, J. Stenarson, D. C. Streit, and H. Zirath, ``Cryogenic wide-band ultra-low-noise IF amplifiers operating at ultra-low DC power,'' \emph{IEEE Trans. Microw. Theory Techn.}, vol. 51, no. 6, pp. 1705--1711, Jun. 2003, doi: 10.1109/TMTT.2003.812570.


\bibitem{Gallego1990}
J. D. Gallego and M. W. Pospieszalski, ``Design and Performance of Cryogenically-Coolable Ultra Low Noise, L-Band Amplifier,'' in \emph{Proc. 20th Eur. Microw. Conf.}, Budapest, Hungary, 1990, pp. 1755--1760, doi: 10.1109/EUMA.1990.336325.

\bibitem{Cano2010} 
J. L. Cano, N. Wadefalk, and J. D. Gallego-Puyol,
``Ultra-wideband chip attenuator for precise noise measurements at cryogenic temperatures,''
\emph{IEEE Transactions on Microwave Theory and Techniques},
vol. 58, no. 9, pp. 2504-2510, 2010.
doi: 10.1109/TMTT.2010.2058276.

\bibitem{Russell2012} 
D. Russell, K. Cleary, and R. Reeves,
``Cryogenic probe station for on-wafer characterization of electrical devices,''
\emph{Review of Scientific Instruments},
vol. 83, no. 4, article number: 044703, 2012.
doi: 10.1063/1.3700213.

\bibitem{Beka2022} 
B. Gabritchidze, K. Cleary, J. Kooi, I. Esho, A. C. Readhead, and A. J. Minnich,
``Experimental characterization of temperature-dependent microwave noise of discrete HEMTs: drain noise and real-space transfer,''
in \emph{2022 IEEE/MTT-S International Microwave Symposium - IMS 2022},
2022, pp. 615-618.
doi: 10.1109/IMS37962.2022.9865505.

\bibitem{Marian1989} 
M. W. Pospieszalski,
``Modeling of noise parameters of MESFETs and MODFETs and their frequency and temperature dependence,''
\emph{IEEE Transactions on Microwave Theory and Techniques},
vol. 37, no. 9, pp. 1340-1350, 1989.
doi: 10.1109/22.32217.

\bibitem{Rorsman1996} 
N. Rorsman, M. Garcia, C. Karlsson, and H. Zirath,
``Accurate small-signal modeling of HFET's for millimeter-wave applications,''
\emph{IEEE Transactions on Microwave Theory and Techniques},
vol. 44, no. 3, pp. 432-437, 1996.
doi: 10.1109/22.486152.

\bibitem{Ardizzi2022} 
A. J. Ardizzi, A. Y. Choi, B. Gabritchidze, J. Kooi, K. A. Cleary, A. C. Readhead, A. J. Minnich,
{“Self-heating of cryogenic high electron-mobility transistor amplifiers and the limits of microwave noise performance”},
\emph{Journal of Applied Physics},
vol. 132, no. 8, pp. 084501, 2022,
doi:10.1063/5.0103156.

\bibitem{Beka2023}
B. Gabritchidze, J. H. Chen, K. A. Cleary, A. C. Readhead, and A. J. Minnich, ``Experimental Investigation of Drain Noise in High Electron Mobility Transistors: Thermal and Hot Electron Noise,'' \emph{IEEE Trans. Electron Devices}, vol. 71, no. 10, pp. 5925--5932, Oct. 2024, doi: 10.1109/TED.2024.3445889.

\bibitem{Murti2000}
M. R. Murti et al., ``Temperature-dependent small-signal and noise parameter measurements and modeling on InP HEMTs,'' \emph{IEEE Trans. Microw. Theory Techn.}, vol. 48, no. 12, pp. 2579--2587, Dec. 2000, doi: 10.1109/22.899016.

\bibitem{Sandy2021} 
S. Weinreb and J. Shi,
{“Low noise amplifier With 7-K noise at 1.4 GHz and 25 °C”},
\emph{IEEE Transactions on Microwave Theory and Techniques},
vol. 69, no. 4, pp. 2345-2351, 2021,
doi:10.1109/TMTT.2021.3061459.

\bibitem{McCulloch2017}
M. A. McCulloch, J. Grahn, S. J. Melhuish, P.-Å. Nilsson, L. Piccirillo, J. Schleeh, and N. Wadefalk, ``Dependence of noise temperature on physical temperature for cryogenic low-noise amplifiers,'' \emph{J. Astron. Telesc. Instrum. Syst.}, vol. 3, no. 1, pp. 014003, 2017, doi: 10.1117/1.JATIS.3.1.014003.

\bibitem{Ohmori2023} 
K. Ohmori and S. Amakawa,
{“Variable-temperature broadband noise characterization of MOSFETs for cryogenic electronics: from room temperature down to 3 K”},
\emph{IEEE Electron Devices Technology \& Manufacturing Conference (EDTM)},
 Seoul, Korea, pp. 1-3, 2023,
doi:10.1109/EDTM55494.2023.10103124.

\bibitem{Schleeh2013}
J. Schleeh, H. Rodilla, N. Wadefalk, P.-Å. Nilsson, and J. Grahn, ``Characterization and Modeling of Cryogenic Ultralow-Noise InP HEMTs,'' \emph{IEEE Trans. Electron Devices}, vol. 60, no. 1, pp. 206--212, Jan. 2013, doi: 10.1109/TED.2012.2227485.

\bibitem{Pospieszalski2017}
M. W. Pospieszalski, ``On the dependence of FET noise model parameters on ambient temperature,'' in \emph{Proc. IEEE Radio Wireless Symp. (RWS)}, Phoenix, AZ, USA, 2017, pp. 159--161, doi: 10.1109/RWS.2017.7885975.


\bibitem{Munoz_1997}
S. Munoz, J. D. Gallego, J. L. Sebastian, and J. M. Miranda, ``Drain Temperature Dependence on Ambient Temperature for a Cryogenic Low Noise C-Band Amplifier,'' in \emph{Proc. 27th Eur. Microw. Conf.}, Jerusalem, Israel, 1997, pp. 114--118, doi: 10.1109/EUMA.1997.337780.

\bibitem{Schleeh2015} 
J. Schleeh, J. Mateos, I. Íñiguez-de-la-Torre, N. Wadefalk, P. A. Nilsson, J. Grahn, and A. J. Minnich,
``Phonon black-body radiation limit for heat dissipation in electronics,''
\emph{Nature materials},
vol. 14, no. 2, pp. 187—192, February 2015.
doi: 10.1038/nmat4126.


\bibitem{Das_2025}
S. Das, S. Raman, and J. C. Bardin, ``Experimental Characterization of the MOSFET Fano Factor at Cryogenic Temperatures for Accurate Cryo-CMOS RF Modeling,'' \emph{IEEE Trans. Microw. Theory Techn.}, vol. 73, no. 10, pp. 7164--7176, Oct. 2025, doi: 10.1109/TMTT.2025.3578317.


\bibitem{Heinz_2022}
F. Heinz, F. Thome, D. Schwantuschke, A. Leuther, and O. Ambacher, ``A Scalable Small-Signal and Noise Model for High-Electron-Mobility Transistors Working Down to Cryogenic Temperatures,'' \emph{IEEE Trans. Microw. Theory Techn.}, vol. 70, no. 2, pp. 1097--1110, Feb. 2022, doi: 10.1109/TMTT.2021.3123647.

\bibitem{Ziel1962}
A. van der Ziel, ``Thermal Noise in Field-Effect Transistors,'' \emph{Proc. IRE}, vol. 50, no. 8, pp. 1808--1812, Aug. 1962, doi: 10.1109/JRPROC.1962.288221.

\bibitem{Hartnagel2001} 
H. L. Hartnagel, R. Katilius, and A. Matulionis,
``Microwave noise in semiconductor devices,''
in \emph{Proceedings of 2001 International Conference on Microwave Noise and Stability},
2001.
Available online: \url{https://api.semanticscholar.org/CorpusID:118611130}.

\bibitem{LiCSW2024}
J. Li, J. Chen, A. J. Minnich, and J. Grahn, ``Investigation of Noise Performance in InP HEMTs with Varying Indium Channel Composition from 80 K to 300 K,'' in \emph{Proc. Compound Semiconductor Week}, Lund, 2024.

\bibitem{Reuter1997} 
R. Reuter, M. Agethen, U. Auer, S. van Waasen, D. Peters, W. Brockerhoff, F.J. Tegude,
{“Investigation and modeling of impact ionization with regard to the RF and noise behavior of HFET”},
\emph{IEEE Transactions on Microwave Theory and Techniques},
vol. 45, no. 6, pp. 977-983, 1997,
doi:10.1109/22.588612.

\bibitem{Mateos1999}
J. Mateos, T. González, D. Pardo, V. Hoel, and A. Cappy, ``Effect of the T-gate on the performance of recessed HEMTs. A Monte Carlo analysis,'' \emph{Semicond. Sci. Technol.}, vol. 14, no. 9, pp. 864, Sep. 1999, doi: 10.1088/0268-1242/14/9/320.

\bibitem{Pearsall1978}
T. P. Pearsall, F. Capasso, R. E. Nahory, M. A. Pollack, and J. R. Chelikowsky, ``The band structure dependence of impact ionization by hot carriers in semiconductors: GaAs,'' \emph{Solid-State Electron.}, vol. 21, no. 1, pp. 297--302, 1978, doi: 10.1016/0038-1101(78)90151-X.

\bibitem{Vurgaftman2001}
I. Vurgaftman, J. R. Meyer, and L. R. Ram-Mohan, ``Band parameters for III--V compound semiconductors and their alloys,'' \emph{J. Appl. Phys.}, vol. 89, no. 11, pp. 5815--5875, Jun. 2001, doi: 10.1063/1.1368156.

\bibitem{Gaquiere1999} 
C. Gaquiere, P. Miraumont, Y. Crosnier, 
{“Measurement technique for determining impact ionisation in HEMTs”},
\emph{Electronics Letters},
vol.35, no.14, pp.1146-1147, 1999,
doi:10.1049/el:19990773.


\bibitem{Zhou2024}
F.-G. Zhou, R.-Z. Feng, S.-R. Cao, Z.-Y. Feng, T. Liu, Y.-B. Su, J.-Y. Shi, W.-C. Ding, and Z. Jin, ``Origin and suppression of kink effect in InP high electron mobility transistors at cryogenic temperatures,'' \emph{Appl. Phys. Lett.}, vol. 3, no. 1, pp. 1--5, 2024, doi: 10.1063/5.0169675.

\bibitem{Vasallo2004} 
B. G. Vasallo, J. Mateos, D. Pardo, T. González,
{“Kink-effect related noise in short-channel InAlAs/InGaAs high electron mobility transistors”},
\emph{Journal of Applied Physics},
vol. 95, no. 12, pp. 8271–8274, 2004,
doi:10.1063/1.1745119.

\bibitem{Huang1992} 
J. H. Huang, T. Y. Chang, B. Lalevic,
{“Measurement of the conduction‐band discontinuity in pseudomorphic InxGa1-xAs/In0.52Al0.48As heterostructures”},
\emph{Applied Physics Letters},
vol. 60, no. 6, pp. 733–735, 1992,
doi.org/10.1063/1.106552.


\bibitem{Zhang2026}
J. Zhang, A. J. Ardizzi, K. A. Cleary, and A. J. Minnich, ``Investigation of Real-Space Transfer Noise in InGaAs/InAlAs Quantum Wells for Indium Phosphide High Electron-Mobility Transistors,'' \emph{Phys. Status Solidi A}, vol. 223, no. 9, pp. e202500981, 2026, doi: 10.1002/pssa.202500981.
















































































\end{thebibliography}
\end{document}